\documentstyle[sprocl]{article}
\input psfig

\newcommand{\nc}{\newcommand}

\nc{\eqr}[1]{(\ref{#1})}
\nc{\sref}[1]{\S\ref{#1}}
\nc{\tref}[1]{Table~\ref{#1}}
\nc{\fref}[1]{Figure~\ref{#1}}
\nc{\cref}[1]{Chapter~\ref{#1}}
\nc{\beq}{\begin{equation}}
\nc{\eeq}{\end{equation}}
\nc{\barray}{\begin{eqnarray}}
\nc{\earray}{\end{eqnarray}}
\nc{\barrayn}{\begin{eqnarray*}}
\nc{\earrayn}{\end{eqnarray*}}
\nc{\bcenter}{\begin{center}}
\nc{\ecenter}{\end{center}}
\nc{\lra}{\longrightarrow}
\nc{\ra}{\rightarrow}
\nc{\setall}{\setcounter{equation}{0}
        \setcounter{definition}{0}
        \setcounter{lemma}{0}
        \setcounter{convention}{0}
        \setcounter{conjecture}{0}
        \setcounter{theorem}{0}
        \setcounter{proposition}{0}
        \setcounter{property}{0}
        \setcounter{fact}{0}
        \setcounter{corollary}{0}}
\nc{\setequation}{\setcounter{equation}{0}}
\nc{\hs}[1]{\hspace{#1 mm}}
\newcommand{\rtimes}{\mbox{$\times\!\rule{0.3pt}{1.1ex}\,$}}

\def\sCC{{\kern 0.27em\vrule height1.45ex width0.03em depth0em
          \kern-0.30em\rm C}}
\def\C{{\mathchoice
  {\sCC}
  {\sCC}
  {\kern 0.225em \vrule height1.05ex width0.025em depth0em \kern-0.25em \rm C}
  {\kern 0.180em \vrule height0.78ex width0.02em depth0em \kern-0.2em \rm C}
        }}
\def\sHH{{\rm I\kern-.16em{}H}}
\def\H{{\mathchoice
  {\sHH}
  {\sHH}
  {\rm I\kern-.13em{}H}
  {\rm I\kern-.13em{}H} }}
\def\sNN{{\rm I\kern-.16em{}N}}
\def\N{{\mathchoice
  {\sNN}
  {\sNN}
  {\rm I\kern-.12em{}N}
  {\rm I\kern-.10em{}N} }}
\def\sPP{{\rm I\kern-.16em{}P}}
\def\P{{\mathchoice
  {\sPP}
  {\sPP}
  {\rm I\kern-.12em{}P}
  {\rm I\kern-.10em{}P} }}
\def\sQQ{{\kern 0.27em \vrule height1.45ex width0.03em depth0em
          \kern-0.30em \rm Q}}
\def\Q{{\mathchoice
        {\sQQ}
        {\sQQ}
  {\kern 0.225em \vrule height1.05ex width0.025em depth0em \kern-0.25em \rm Q}
  {\kern 0.180em \vrule height0.78ex width0.020em depth0em \kern-0.20em \rm Q}
        }}
\def\sRR{{\rm I\kern-0.16em{}R}}
\def\R{{\mathchoice
  {\sRR}
  {\sRR}
  {\rm I\kern-0.12em{}R}
  {\rm I\kern-0.10em{}R} }}
\def\sZZ{{\rm Z\kern-0.32em{}Z}}
\def\Z{{\mathchoice
  {\sZZ}
  {\sZZ} 
  {\rm Z\kern-0.3em{}Z}     
  {\rm Z\kern-0.25em{}Z} }}  
\def\ZZZ{{\rm Z\kern-0.24em{}Z}}
\def\sKK{{\rm I\kern-0.16em{}K}}
\def\K{{\mathchoice
  {\sKK}
  {\sKK}
  {\rm I\kern-0.12em{}K}
  {\rm I\kern-0.10em{}K} }}
\def\odd{{\rm odd}}
\def\even{{\rm even}}
\def\gcd{{\rm gcd}}

\newtheorem{proposition}{\bf PROPOSITION}

\begin{document}

{\flushright{\small MIT-CTP-2897\\}}

\title{$Z$-$D$ Brane Box Models and Non-Chiral Dihedral Quivers.}

\author{Bo Feng, Amihay Hanany and Yang-Hui He}
\address{fengb, hanany, yhe@ctp.mit.edu\footnote{
Research supported in part
by the CTP and the LNS of MIT and the U.S. Department of Energy 
under cooperative research agreement $\#$DE-FC02-94ER40818; YHH is also
supported by the NSF Graduate Fellowship.}
\\}
\address{\it Center for Theoretical Physics,\\ Massachusetts
Institute of Technology\\ Cambridge, Massachusetts 02139, U.S.A.\\}

\maketitle\abstracts{Generalising ideas of an earlier work \cite{Bo-Han}, 
we address the problem of
constructing Brane Box
Models of what we call the $Z$-$D$ Type from a new point of view, so as
to establish the complete correspondence between these brane setups
and orbifold singularities of the non-Abelian
$G$ generated by $Z_k$ and $D_d$ under certain 
group-theoretic constraints to
which we refer as the BBM conditions.
Moreover, we present
a new class of ${\cal N}=1$ quiver theories of the ordinary 
dihedral group $d_k$ as well as the ordinary exceptionals $E_{6,7,8}$
which have
non-chiral matter content and discuss issues related to
brane setups thereof.}

\section{Introduction}
Configurations of branes \cite{Han-Wit} have been proven to be a 
very useful method to
study the gauge field theory which emerges as the low energy limit of string
theory (for a complete reference, see Giveon and Kutasov \cite{Giveon}).
The advantage of 
such setups is that they provide an intuitive picture so that we can very easily
deduce many properties of the gauge theory.
For example, brane setups have been used to study mirror symmetry in 3 
dimensions \cite{Han-Wit,IS,P-Zaf,Boer,Kapustin}, Seiberg Duality in 4
dimensions \cite{Elitzur},
and exact solutions when lifting Type IIA setups to M-theory \cite{Mlift,Karl}.
After proper T- or S-dualities, we can
transform the above brane setups to D-brane as probes on some target space with
orbifold singularities \cite{Quiver,Johnson,Law-Vafa}.

For example, the brane setup
of streching Type IIA D4-branes between $n+1$ NS5-branes placed in a circular
fashion (the ``elliptic model'' \cite{Mlift}) is precisely T-dual to D3-branes
stacked upon ALE singularities of the type $\widehat{A_n}$, or in other words 
orbifold singularities of the form $\C^2/Z_{n+1}$,  where $Z_{n+1}$ is
the cyclic group on $n+1$ elements and is a finite discrete 
subgroup of $SU(2)$.
As another example, the Brane Box Model \cite{Han-Zaf,Han-Ura,Han-S}
is T-dual to D3-branes as probes on orbifold singularities of the
type $\C^3/\Gamma$ with $\Gamma=Z_k$ or $Z_k \times Z_{k'}$ now being a finite
discrete subgroup of $SU(3)$ \cite{Han-Ura}. A brief review of some of these 
contemporary techniques can be found in our recent work \cite{Bo-Han}.
In fact, it is a very interesting problem to see how in general 
the two different methods, viz., brane setups and D3-branes
as probes on geometrical singularities, are connected to each other
by proper duality transformations \cite{Karch1}.

The general construction and methodology for D3-branes as probes
on orbifold singularities has been given \cite{Law-Vafa}. However, the complete
list of the
corresponding brane setups is not yet fully known. For orbifolds 
$\C^2/\{\Gamma \in SU(2)\}$, we have the answer
for the $\widehat{A_n}$ series (i.e., $\Gamma=Z_{n+1}$) and the 
$\widehat{D_n}$ series (i.e., $\Gamma=D_{n-2}$, the binary dihedral groups) 
\cite{Kapustin}, but not for the exceptional cases $\widehat{E_{6,7,8}}$.
At higher dimensions, the situation is even more disappointing: for orbifolds of 
$\C^3/\{\Gamma \in SU(3)\}$, brane setups are until recently limited to only
Abelian singularities, namely $\Gamma=Z_k$ or $Z_{k} \times Z_{k'}$ \cite{Han-Ura}.

In a previous paper \cite{Bo-Han}, we went beyond the Abelian restriction in
three dimensions and gave 
a new result concerning the correspondence of the two methods.
Indeed we showed that\footnote{In that paper we used the notation
$Z_k \times D_{k'}$ and pointed out that the symbol $\times$ was
really an abuse. We shall here use the symbol $*$ and throughout the
paper reserve $\times$ to mean strict direct product of groups and
$\rtimes$, the semi-direct product.} for $\Gamma := G = Z_k * D_{k'}$ 
a finite 
discrete subgroup of $SU(3)$, the corresponding brane setup (a Brane Box Model)
T-dual to the orbifold discription can be obtained.
More explicitly, the group $G \in SU(3)$ is defined as being generated by the
following matrices that act on $\C^3$:

\begin{equation}
\label{gen1}
\alpha = \left(  \begin{array}{ccc}
                        \omega_{k} & 0 & 0  \\
                         0 & \omega_{k}^{-1} & 0\\
                         0  &  0 & 1
                \end{array}
        \right)
~~~~~~~~
\beta = \left(  \begin{array}{ccc}
                        1  & 0  & 0  \\
                        0 & \omega_{2k'} & 0  \\
                        0 &  0 & \omega_{2k'}^{-1}  
                \end{array}
        \right)
~~~~~~~~
\gamma =\left(  \begin{array}{ccc} 
                1  &  0  &  0  \\
                0  &  0  &  i \\
                0  &  i &   0 
                \end{array}
        \right)
\end{equation}
where $\omega_n := e^{\frac{2 \pi i}{n}}$, the $n$th root of unity.

The abstract presentation of the groups is as follows:
\begin{equation}
\alpha \beta = \beta \alpha,~~~~\beta \gamma =\gamma \beta^{-1},~~~~
\alpha^{m} \gamma \alpha^{n} \gamma =\gamma \alpha^{n} \gamma \alpha^{m}
~~~~\forall m,n \in \Z
\label{rel1}
\end{equation}

Because of the non-Abelian property of $G$,
the preliminary attempts at the corresponding Brane Box Model by using the idea
in a previous work \cite{Han-Zaf2} met great difficulty.
However, via careful analysis, 
we found that the group $G$ can be written as the semidirect product of $Z_{k}$ and
$D_{\frac{kk'}{\gcd(k,2k')}}$. Furthermore, when $\frac{2k'}{\gcd(k,2k')}=\even$,
the character table of $G$ as the semidirect product 
$Z_k \rtimes D_{\frac{kk'}{\gcd(k,2k')}}$
preserves the structure of that of $D_{\frac{kk'}{\gcd(k,2k')}}$,
in the sense that it seems to be composed of $k$ copies of the latter.
Indeed, it was noted \cite{Bo-Han} that only under this parity condition 
of $\frac{2k'}{\gcd(k,2k')}=\even$, can we construct, with
the two group factors $Z_{k}$ and $D_{\frac{kk'}{\gcd(k,2k')}}$, a consistent
Brane Box Model with the ideas in the abovementioned paper \cite{Han-Zaf2}.

The success of the above construction, constrained by certain 
conditions, hints that
something fundamental is acting as a key r\^ole in the construction of
non-Abelian brane setups above two (complex) dimensions.
By careful consideration, it seems that the following three conditions
presented themselves to be crucial in the study of $Z_k * D_{k'}$
which we here summarize:
\begin{enumerate}
\label{condition}
\item The whole group $G$ can be written as a semidirect product:
	$Z_k \rtimes D_{d}$;

\item The semidirect product of $G$ preserves the structure of the 
	irreducible representations of $D_{d}$, i.e., it appears
	that the irreps of $G$
	consist of $k$ copies of those of the subgroup $D_{d}$;

\item There exists a (possibly reducible) representation of $G$ 
	in 3 dimensions such that the representation matrices belong to $SU(3)$.
	Henceforth, we shall call such a representation,
	consistent with the $SU(3)$ requirement (see more discussions
	\cite{Bo-Han,Han-He} on decompositions), as 
	``{\it the chosen decomposition of {\bf 3}}''.
\end{enumerate}
We will show in this paper that these conditions are sufficient for
constructing Brane Box Model of the $Z$-$D$ type. Here we will call the
Brane Box Model in our recent paper \cite{Bo-Han} as 
{\bf Type $Z$-$D$} and similarly, 
that in earlier works \cite{Han-Zaf,Han-Ura} we shall call the
$Z$-$Z$ Type. We shall see this definition more clearly in subsection
\sref{subsec:BBMZD}.
It is amusing to notice that Brane Box Models of Type $Z$-$Z$ also 
satisfy the above three conditions since they correspond to the
group $Z_k \times Z_{k'}$, which is a direct product.

Furthermore, we shall answer a mysterious question posed
at the end of our earlier work \cite{Bo-Han}. 
In that paper, we discussed the so-called
{\it Inverse Problem}, i.e., given a consistent Brane Box Model, how
may one determine, from the structure of the setup 
(the number and the positioning
of the branes), the corresponding group $\Gamma$ in the orbifold 
structure of $\C^3/\Gamma$.
We found there that only when $k$ is the divisor of $d$ can we
find the corresponding group defined in (\ref{gen1}) with proper
$k,k'$. This was very unsatisfying. However, the structure of the
Brane Box Model of Type $Z$-$D$ was highly suggestive of the solution for
general $k,d$. We shall here mend that short-coming and for
arbitrary $k,d$ we shall construct the
corresponding group $\Gamma$ which satisfies above three conditions. 
With this result, we establish the complete correspondence 
between the Brane Box Model of Type $Z$-$D$ and 
D3-branes as probes on orbifold singularities
of $\C^3/\Gamma$ with properly determined $\Gamma$.

The three conditions which are used for solving the inverse problem can be 
divided into two conceptual levels. The first two are at the level of pure
mathematics, i.e., we can consider it from the point of view of abstract group theory
without reference to representations or to finite discrete subgroup of
$SU(n)$. The third condition is at the level of physical applications. From
the general structure \cite{Law-Vafa} we see that for constructing 
${\cal N}=2$
or ${\cal N}=1$ theories we respectively need the group $\Gamma$ to be a finite 
subgroup of $SU(2)$ or $SU(3)$. This requirement subsequently means that
we can find a faithful (but possibly reducible) 2-dimensional or 3-dimensional 
representation with the matrices satisfying the determinant 1 and unitarity
conditions. 
In other words, what supersymmetry (${\cal N}=2$ or 1) we 
will have in the orbifold theory by the standard procedure 
\cite{Law-Vafa} depends only on the chosen representation (i.e., the
decomposition of ${\bf 2}$ or ${\bf 3}$). Such distinctions were already
shown before \cite{Han-Ura,Han-He}. The 
group $Z_3$ had been considered \cite{Han-Ura}.
If we choose its action on $\C^3$ as 
$(z_1,z_2,z_3) \longrightarrow
(e^{\frac{2\pi i}{3}} z_1,e^{\frac{-2\pi i}{3}} z_2,z_3)$
we will have ${\cal N}=2$ supersymmetry, but if we choose
the action to be 
$(z_1,z_2,z_3)\longrightarrow
(e^{\frac{2\pi i}{3}} z_1,e^{\frac{2\pi i}{3}} z_2,e^{\frac{2\pi i}{3}} z_3)$
we have only ${\cal N}=1$. This phenomenon mathematically corresponds to 
what are called sets of transitivity of collineation group actions \cite{Trans}.

Moreover, we notice that the ordinary dihedral group $d_k$ which is
excluded from the classification of finite subgroup of $SU(2)$ can be
imbedded\footnote{Since it is in fact a subgroup of $SU(2)/\Z_2 \cong SO(3)$,
the embedding is naturally induced from $SO(3) \hookrightarrow SU(3)$.
In fact the 3-dimensional representation in $SU(3)$ is faithful.} into $SU(3)$.
Therefore we expect that $d_k$ should be useful in constructing some
${\cal N}=1$ gauge field theories by the standard 
procedures \cite{Law-Vafa,Han-He}.
We show in this paper that this is so.
With the proper decompositions, we can obtain new types of gauge theories by 
choosing $\C^3$ orbifolds to be of the type $d_k$.
For completeness, we also give the quiver 
diagrams of ordinary tetrahedral, octahedral and icosahedral groups 
($E_{6,7,8}$), which by a similar token, can be imbedded into $SU(3)$.

The organisation of the paper is as follows. In \sref{sec:dir} we give a simple
and illustrative example of constructing a Brane Box Model for the direct
product $Z_k \times D_{k'}$, whereby initiating the study of brane setups
of what we call Type $Z$-$D$. In \sref{sec:twist} we deal
with the twisted case which we encountered earlier \cite{Bo-Han}.
We show that we can imbed the latter into the direct product (untwisted) case of
\sref{sec:dir} and arrive at another member of Brane Box Models of the $Z$-$D$ 
type. 
In \sref{sec:new} we give a new class of $SU(3)$ quiver
which are connected to the ordinary dihedral group $d_k$. Also, we give 
an interesting brane configuration that will give matter matter content as the 
$d_{k=\even}$ quiver but a different superpotential on the gauge theory level.
Finally in \sref{sec:con} we give concluding remarks and suggest future prospects.

\section*{Nomenclature}
Unless otherwise specified, we shall throughout the paper adhere to the
notation that the group binary operator $\times$ refers to the 
strict direct product,
$\rtimes$, the semi-direct product, and $*$, a general (twisted)
product by combining the generators of the 
two operands\footnote{Therefore in the previous paper \cite{Bo-Han}, the
group $G := Z_k \times D_{k'}$ in this convention should be written as
$Z_k*D_{k'}$, {\it q.v. Ibid.} for discussions on how these different group
compositions affect brane constructions.}.
Furthermore, $\omega_n$ is defined to be $e^{\frac{2 \pi i}{n}}$, the $n$th
root of unity; $H \triangleleft G$ mean that $H$ is a normal subgroup of $G$; and
a group generated by the set $\{x_i\}$ under relations 
$f_i(\{x_j\}) = 1$ is denoted as $\langle x_i | f_j \rangle$.

\section{A Simple Example: The Direct Product $Z_{k} \times D_{k'}$}
\label{sec:dir}
We recall that in a preceeding work \cite{Bo-Han}, 
we constructed the Brane Box Model (BBM) for
the group $Z_k * D_{k'}$ as generated by (\ref{gen1}), satifying 
the three conditions mentioned above, which we shall dub as the 
{\bf BBM condition}
for groups. However, as we argued in the introduction, there may exist in
general, groups not isomorphic to the one addressed \cite{Bo-Han} but still
obey these conditions.
As an illustrative example, we start with the simplest member of the family of
$Z*D$ groups that satisfies the BBM condition, namely the
direct product $G=Z_k \times D_{k'}$. We define $\alpha$
as the generator for the $Z_{k}$ factor and $\gamma,\beta$, those for
the $D_{k'}$. Of course by definition of the direct product $\alpha$ must
commute with both $\beta$ and $\gamma$.
The presentation of the group is clearly as follows:
\[
\begin{array}{ll}
\alpha^k=1; & 	$The Cyclic Group $Z_k \\
\beta^{2k'} =1,~~~~~\beta^{k'}=\gamma^2,~~~\beta \gamma=\gamma \beta^{-1}; &
	$The Binary Dihedral Group $D_{k'} \\
\alpha \beta =\beta \alpha,~~~\alpha \gamma =\gamma \alpha &
	$Mutual commutation$ \\
\end{array}
\]

We see that the first two of the BBM conditions are trivially satisfied.
To satisfy the third, we need a 3-dimensional matrix 
represenation of the group.
More explicitly, as discussed \cite{Bo-Han}, to construct
the BBM of the $Z$-$D$ type, one needs the decomposition of ${\bf 3}$
into one nontrivial 1-dimensional irrep and one 2-dimensional 
irrep. In light of this, we can write down the $SU(3)$ matrix
generators of the group as

\begin{equation}
\label{gen_dir}
\alpha = \left(  \begin{array}{ccc}
			\omega_{k}^{2} & 0 & 0  \\
			 0 & \omega_{k}^{-1} & 0\\
			 0  &  0 & \omega_{k}^{-1}
		\end{array}
	\right)
~~~~~~~~
\beta = \left(  \begin{array}{ccc}
			1  & 0  & 0  \\
			0 & \omega_{2k'} & 0  \\
			0 &  0 & \omega_{2k'}^{-1}  
		\end{array}
	\right)
~~~~~~~
\gamma =\left(  \begin{array}{ccc} 
		1  &  0  &  0  \\
		0  &  0  &  i \\
		0  &  i &   0 
		\end{array}
	\right)
\end{equation}

Here, we notice a subtle point. When $k=\even$, $\alpha^{\frac{k}{2}}$
and $\beta^{k'}$ give the same matrix form. In other words,
(\ref{gen_dir}) generates a {\it non-faithful} representation. We will
come back to this problem later, but before diving into a detailed 
discussion on the whole group $Z_k \times D_{k'}$, let us first give 
the necessary properties of the factor $D_{k'}$.

\subsection{The Group $D_{k'}$}
\label{subsec:Dk'}
One can easily check that all the elements of the binary dihedral
$D_{k'}=\langle \beta, \gamma \rangle$ group can be written, 
because $\gamma^2=\beta^{k'}$, as
\[
\gamma^n \beta^p,~~~{\rm with}~~~n=0,1~~p=0,1,...,2k'-1.
\]

From this constraint and the conjugation relation 
\[
(\gamma^{n_1} \beta^{p_1})^{-1} (\gamma^{n} \beta^{p}) 
(\gamma^{n_1} \beta^{p_1}) = 
\gamma^{n} \beta^{p_1(1-(-1)^{n})+(-1)^{n_1}p},
\]
we can see that the group is of order $4k'$ and moreover affords
4 1-dimensional irreps and $(k'-1)$ 2-dimensional irreps. The
classes of the group are:
\[
\begin{array}{ccccc}
        & C_{n=0}^{p=0} & C_{n=0}^{p=k'} & C_{n=0}^{\pm p} & C_{n=1}^{p \bmod 2} \\
|C|  & 1 & 1 & 2 & k' \\
\#C   & 1 & 1 & k'-1 & 2
\end{array}
\]

To study the character theory of $G := D_{k'}$, we recognise that
$H := \{\beta^{p}\}$ for $p$ even is a normal subgroup of $G$. Whence we can use
the Frobenius-Clifford theory of induced characters
to obtain the irreps of $G$ from the
factor group $\widetilde{G} :=G/H={1,\beta,\gamma,\gamma \beta}$.
For $k'$ even, $\widetilde{G}$ is $Z_2\times Z_2$ and for $k'$ odd, it is simply $Z_4$.
these then furnish the 1-dimensional irreps.
We summarise the characters of these 4 one dimensionals as follows:
\[
\begin{array}{c|c}
k' = \even
&
k' = \odd \\ \hline
\begin{array}{c|cccc}
     &	\beta^{p=\even} & \beta(\beta^{\odd}) & \gamma(\gamma \beta^{\even}) &
	\gamma \beta (\gamma \beta^{\odd}) \\
\chi^{1} & 1 & 1 & 1 & 1 \\
\chi^{2} & 1 & -1 & 1 & -1 \\
\chi^{3} & 1 & 1 & -1 & -1 \\
\chi^{4} & 1 & -1 & -1 & 1 
\end{array}
&
\begin{array}{cccc}
	\beta^{\even} & \beta(\beta^{\odd}) & \gamma(\gamma \beta^{\even}) &
	\gamma \beta (\gamma \beta^{\odd}) \\
1 & 1 & 1 & 1 \\
1 & -1 & \omega_4 & -\omega_4 \\
1 & 1 & -1 & -1 \\
1 & -1 & - \omega_4 &  \omega_4
\end{array} \\
\end{array}
\]

The 2-dimensional irreps can be directly obtained from the
definitions; they are indexed by a single integer $l$:

\begin{equation}
\label{2dreps}
\chi_{2}^{l}(C_{n=1})=0,~~~~\chi_{2}^{l}(C_{n=0}^{p})= 
(\omega_{2k'}^{lp}+\omega_{2k'}^{-lp}),~~l=1,..,k'-1.
\end{equation}

The matrix representations of these 2-dimensionals are given below:
\[
\beta^{p} = \left(  \begin{array}{cc}  \omega_{2k'}^{lp} & 0 \\
					0 & \omega_{2k'}^{-lp}
			\end{array}
		\right)
~~~~~~~
\gamma \beta^{p} = \left(  \begin{array}{cc}  0 & i^l\omega_{2k'}^{-lp} \\
					i^l \omega_{2k'}^{lp} & 0 
			\end{array}
		\right)
\]
 From (\ref{2dreps}) we immediately see that 
$\chi_2^{l}=\chi_2^{-l}=\chi_2^{2k'-l}$ which we use to restrict the index
$l$ in $\chi_2^l$ into the region $[1,k'-1]$. 

Now for the purposes of the construction of the BBM, we aboveall need to 
know the tensor decompositions of the group; these we summarise below.
\[
\begin{array}{|l|l|}
\hline
{\bf 1} \otimes {\bf 1}'
&
\begin{array}{c|c}
	k' = \even	& k' = \odd \\
	\begin{array}{ccc}
	\chi_1^2\chi_1^2=\chi_1^1  & \chi_1^3\chi_1^3=\chi_1^1 &
	\chi_1^4\chi_1^4=\chi_1^1  \\
	\chi_1^2 \chi_1^3=\chi_1^4 & \chi_1^2\chi_1^4=\chi_1^3 &
	\chi_1^3\chi_1^4=\chi_1^2
	\end{array}
	&
	\begin{array}{ccc}
	\chi_1^2\chi_1^2=\chi_1^3  & \chi_1^3\chi_1^3=\chi_1^1 &
	\chi_1^4\chi_1^4=\chi_1^3  \\
	\chi_1^2 \chi_1^3=\chi_1^4 & \chi_1^2\chi_1^4=\chi_1^1 &
	\chi_1^3\chi_1^4=\chi_1^2
	\end{array}
\end{array}
\\ \hline
{\bf 1} \otimes {\bf 2}
&
\chi_1^{h} \chi_2^l = \left\{ \begin{array}{l}
\chi_2^l~~~~h=1,3  \\
\chi_2^{k'-l}~~~~h=2,4 
\end{array}
\right.
\\ \hline
{\bf 2} \otimes {\bf 2'}
&
\chi_2^{l_1} \chi_2^{l_2}=\chi_2^{(l_1+l_2)}+\chi_2^{(l_1-l_2)}
{\rm ~where~}
\begin{array}{l}
	\chi_2^{(l_1+l_2)}= \left\{ \begin{array}{l}
	\chi_2^{(l_1+l_2)}~~~~{\rm if}~~~l_1+l_2<k',  \\
	\chi_2^{2k'-(l_1+l_2)}~~~~{\rm if}~~~l_1+l_2>k', \\
	\chi_1^2+\chi_1^4~~~~{\rm if}~~~l_1+l_2=k'.
	\end{array}
	\right.
	\\
	\chi_2^{(l_1-l_2)}= \left\{ \begin{array}{l}
	\chi_2^{(l_1-l_2)}~~~~{\rm if}~~~l_1>l_2,  \\
	\chi_2^{(l_2-l_1)}~~~~{\rm if}~~~l_1<l_2, \\
	\chi_1^1+\chi_1^3~~~~{\rm if}~~~l_1=l_2.
	\end{array}
	\right.
\end{array}
\\
\hline
\end{array}
\]

\subsection{The Quiver Diagram}
The general method of constructing gauge field theories from orbifold 
singularities of $C^3/\Gamma \subset SU(3)$ has been given \cite{Law-Vafa,Han-He}.
Let us first review briefly the essential results.
Given a finite discrete subgroup $\Gamma \subset SU(3)$ with irreducible
representations $\{ r_i \}$,
we obtain, under the orbifold projection, an ${\cal N}=1$ super Yang-Mills 
theory with gauge group
\[
\bigotimes_{i} SU(N|r_i|),~~~~~~|r_i|=\dim(r_i),N \in \Z
\]
To determine the matter content we need to choose the decomposition of ${\bf 3}$
(i.e., the $3 \times 3$ matrix form) of $\Gamma$ which describes how it acts upon
$\C^3$. 
We use $R$ to denote the representation of chosen ${\bf 3}$ and
calculate the tensor decomposition 

\begin{equation}
\label{decomp}
R \otimes r_i= \bigoplus_{j} a_{ij}^{R} r_{j}
\end{equation}

The matrix $a_{ij}^{R}$ then tells us how many bifundamental chiral multiplets
of $SU(N_i) \times SU(N_j)$ there are which transform under the representation 
$(N_i,\bar{N_j})$, where $N_i := N|r_i|$.
Furthermore, knowing this matter content we can also write down
the superpotential whose explicit form is given in (2.7) and (2.8) of
Lawrence, Nekrasov and Vafa \cite{Law-Vafa}.
We do not need the detailed form thereof but we emphasize that all terms in 
the superpotential are cubic and there are no quatic term. This condition
is necessary for finiteness \cite{Han-S,Law-Vafa}
and we will turn to this fact later.

We can encode the above information into a ``{\bf quiver diagram}''.
Every node $i$ with index $\dim{r_i}$ in the 
quiver denotes the gauge group $SU(N_i)$. 
Then we connect $a_{ij}^{R}$ arrows from node $i$ to $j$ in order
to denote the correpsonding bifundamental chiral multiplet $(N_i,\bar{N_j})$.
When we say that a BBM construction is {\bf consistent} we mean that
it gives the same quiver diagram as one obtains from the geometrical
probe methods \cite{Law-Vafa}.

Now going back to our example $Z_k \times D_{k'}$, its
character table is easily written: it is simply the 
Kronecker product of the
character tables of $Z_k$ and $D_{k'}$ (as matrices).
We promote (\ref{2dreps}) to a double index
\[
(a,\chi_{i}^{l})
\]
to denote the charaters, where $a = 0, ... , k-1$ and
are characters of $Z_k$ (which are simply various $k$th roots of unity)
and $\chi$ are the characters of $D_{k'}$ as presented in the previous
subsection. We recall that $i = 1$ or 2 and for the former, there
are 4 1-dimensional irreps indexed by $l=1,..,4$; and for the latter, 
there are $k'-1$ 2-dimensional irreps indexed by $l=1,..,k'-1$.
It is not difficult to see from (\ref{gen_dir}) that the chosen decomposition 
should be:
\[
{\bf 3} \longrightarrow (2,\chi_{1}^{1}) \oplus (-1,\chi_{2}^{1})
\]
The relevant tensor decomposition which gives the quiver is then
\begin{equation}
\label{tensor}
[(2,\chi_{1}^{1}) \oplus (-1,\chi_{2}^{1})]\otimes (a,\chi_{i}^{l})
=(a+2,\chi_{i}^{l}) \oplus (a-1,\chi_{i}^{l} \otimes \chi_{2}^{1}),
\end{equation}
which is thus reduced to the decompositions as tabulated in the previous
subsection.

\subsection{The Brane Box Model of $Z_k \times D_{k'}$}
\label{subsec:BBMZD}
Now we can use the standard methodology \cite{Han-Ura,Bo-Han,Han-Zaf2}
to construct the BBM. The general 
idea is that for the BBM corresponding to the singularity $\C^3/\Gamma$,
we put D-branes whose number is determined by the irreps of $\Gamma$ 
into proper grids in Brane Boxes constructed out of NS5-branes.
Then the genetal rule of the resulting BBM is that we have gauge group
$SU(N_i)$ in every grid and bifundamental chiral multiplets going along
the North, East and SouthWest directions.
The superpotential can also
be read by closing upper or lower triangles in the grids \cite{Han-Ura}.
The quiver diagram is also readily readable from the structure of the
BBM (the number and the positions of the branes).

Indeed, in comparison with geometrical methods, because the two quivers
(the orbifold quiver and the BBM quiver) seem to arise from two 
vastly disparate contexts, they need not match a priori.
However, by judicious choice of irreps in each grid
we can make these two quiver exactly the same; this is what is meant
by the {\bf equivalence} between the BBM and orbifold methods.
The consistency condition we impose on the BBM for such equivalence is

\begin{equation}
\label{consistency}
{\bf 3} \otimes r_i = \bigoplus_{j \in \{{\rm North,East,SouthWest}\}} r_j.
\end{equation}

Of course we observe this to be precisely (\ref{decomp}) in a different
guise.

\begin{figure}
\centerline{\psfig{figure=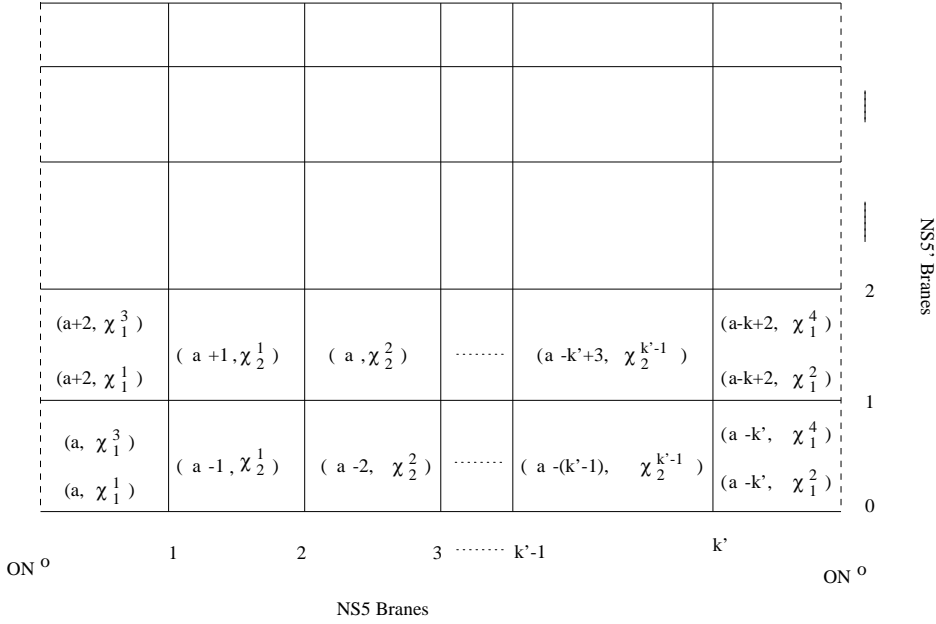,width=5.0in}}
\caption{The Brane Box Model for $Z_k \times D_{k'}$. Notice that 
	for every step along
	the vertical direction from the bottom to top, 
	the first index has increment 2, while along the 
	horizontal direction from left to right, 
	the first index has decrement 1 and
	the second index, increment 1. The vertical direction
	is also periodically identified so that $k(~$or$~\frac{k}{2}) \equiv 0$.}
\label{fig:BBM1}
\end{figure}

Now we return to our toy group $Z_k \times D_{k'}$. The grids are furnished
by a parallel set of $k'$ NS5-branes with 2 ON$^0$ planes intersected by
$k$ (or $\frac{k}{2}$ when $k$ is even; see explanation below) 
NS5$'$-branes perpendicular thereto and periodically identified such
that $k($or$~\frac{k}{2}) \equiv 0$ as before \cite{Bo-Han}. This is shown in 
\fref{fig:BBM1}. The general brane setup of this form involving
2 sets of NS5-branes and 2 ON-planes we shall call, as mentioned in
the introduction, the BBM of {\bf the $Z$-$D$ Type}.

The irreps are placed in the grids as follows.
First we consider the leftmost column.
We place a pair of  irreps $\{(a,\chi_1^1),(a,\chi_1^3)\}$ at the bottom
(here $a$ is some constant initial index), 
then for each incremental grid going up we increase the index $a$ by 2.
Now we notice the fact that when $k$ is odd, such an indexing makes one
return to the bottom grid after $k$ steps whereas if $k$ is even, it suffices
to only make $\frac{k}{2}$ steps before one returns.
This means that when $k$ is odd, the periodicity of $a$ is
precisely the same as that required by our circular
identification of the NS5$'$-branes. However, when $k$ is even it seems
that we can not put all irreps into a single BBM. We can circumvent the
problem by dividing the irreps $(a,\chi)$ into 2 classes depending on
the parity of $a$, each of which gives a BBM consisting of $k\over 2$
NS5$'$-branes.
We should not be surprised at this phenomenon. As we mentioned at
the beginning of this section, the matrices (\ref{gen_dir}) generate a
non-faithful representation of the group when $k$ is even
(i.e., $\alpha^{\frac{k}{2}}$ gives the same matrix as $\beta^{k'}$).
This non-faithful decomposition of ${\bf 3}$ is what is responsible for
breaking the BBM into 2 disjunct parts.

The same phenomenon appears in the $Z_k \times Z_{k'}$ BBM as well.
For $k$ even, if we choose the decomposition as 
${\bf 3} \longrightarrow(1,0)+(0,1)+(-1,-1)$ 
we can put all irreps into $kk'$ grids, however
if we choose
${\bf 3} \longrightarrow (2,0)+(0,1)+(-2,-1)$
we can only construct two BBM's each with $\frac{kk'}{2}$ grids and
consisting of one half of the total irreps.
Indeed this a general phenomenon which we shall use later:

\begin{proposition}
\label{nonfaith}
Non-faithful matrix representations of $\Gamma$ give rise to
corresponding Quiver Graphs which are disconnected.
\end{proposition}

Having clarified this subtlety we continue to construct the BBM.
We have fixed the content for the leftmost column.
Now we turn to the bottom row.
Starting from the second column (counting from the
left side) we place the irreps
$(a-1,\chi_2^{1}), (a-2,\chi_2^{2}) ,..., (a-(k'-1),\chi_2^{k'-1})$
until we reach the right side (i.e., $(a-j,\chi_2^{j})$ with $j=1,...k'-1$)
just prior to the rightmost column; there we place
the pair $\{(a-k',\chi_1^2),(a-k',\chi_1^4)\}$.
For the remaining rows we imitate what we did for the leftmost column
and increment, for the $i$-th column, the first index by 2 each time
we ascend one row, i.e., $(b,\chi_i^{j}) \rightarrow (b+2,\chi_i^{j})$.
The periodicity issues are as discussed above.

Our task now is to check the consistency of the BBM, namely 
(\ref{consistency}).
Let us do so case by case. First we check the grid at the first 
(leftmost) column at the $i$-th row; the content there is
$\{(a+2i,\chi_1^1),(a+2i,\chi_1^3)\}$.
Then (\ref{consistency}) dictates that
\[
\begin{array}{c}
[(2,\chi_{1}^{1}) \oplus (-1,\chi_{2}^{1})] 
	\otimes(a+2i,\chi_1^1 ~$or$~ \chi_1^3) \\
=(a+2(i+1),\chi_1^1 ~$or$~ \chi_1^3) \oplus ((a+2i)-1,\chi_{2}^{1})
\end{array}
\]
by using the table of tensor decompositions in subsection \sref{subsec:Dk'}
and our chosen {\bf 3} from (\ref{tensor}). Notice that the
first term on the right is exactly the content of the box to the
North and second term, the content of the East. Therefore consistency
is satisfied.
Next we check the grid in the second column at the $i$-th row where
$((a+2i)-1,\chi_{2}^{1})$ lives. As above we require
\[
\begin{array}{c}
[(2,\chi_{1}^{1}) \oplus (-1,\chi_{2}^{1})]\otimes ((a+2i)-1,\chi_{2}^{1}) \\
=((a+2(i+1))-1,\chi_{2}^{1}) \oplus ((a+2i)-2,\chi_{2}^{2})
	\oplus (a+2(i-1),\chi_1^1) \oplus (a+2(i-1),\chi_1^3)
\end{array}
\]
whence we see that the first term corresponds to the grid to the North, 
and second, East, and the last two, SouthWest. 
We proceed to check the grid in the $j+1$-th column ($2\leq j\leq k'-2$) at
the $i$-th row where $((a+2i)-j,\chi_2^{j})$ resides.
Again (\ref{consistency}) requires that
\[
\begin{array}{c}
[(2,\chi_{1}^{1}) \oplus (-1,\chi_{2}^{1})]\otimes ((a+2i)-j,\chi_2^{j}) \\
=((a+2(i+1))-j,\chi_2^{j}) \oplus ((a+2i)-(j+1),\chi_2^{j+1})
	\oplus ((a+2(i-1))-(j-1),\chi_2^{j-1})
\end{array}
\]
where again the first term gives the irrep the grid to the North, 
the second, East and the third, SouthWest. 
Next we check the grid in the $k'$-th column and $i$-th row, where the
irrep is $((a+2i)-(k'-1),\chi_2^{k'-1})$. Likewise the requirement is
\[
\begin{array}{c}
[(2,\chi_{1}^{1}) \oplus (-1,\chi_{2}^{1})]\otimes ((a+2i)-(k'-1),\chi_2^{k'-1})\\
=((a+2(i+1))-(k'-1),\chi_2^{k'-1}) \oplus ((a+2i)-k',\chi_1^{2}) \\ \oplus
	((a+2i)-k',\chi_1^{4}) \oplus ((a+2(i-1))-(k'-2),\chi_2^{k'-2})
\end{array}
\]
whence we see again the first term gives the grid to the North, the second 
and third, East and the fourth, SouthWest. 
Finally, for the last (rightmost) column, the grid in the $i$-th row has
$((a+2i)-k',\chi_1^{2})$ and $((a+2i)-k',\chi_1^{4})$. We demand
\[
\begin{array}{c}
[(2,\chi_{1}^{1}) \oplus (-1,\chi_{2}^{1})]\otimes 
	((a+2i)-k',\chi_1^{2} ~$or$~ \chi_1^{4}) \\
=((a+2(i+1))-k',\chi_1^{2} ~$or$~ \chi_1^{4}) \oplus 
	((a+2(i-1))-(k'-1),\chi_{2}^{k'-1}))
\end{array}
\]
where the first term gives the grid to the North and the second term,
Southwest. So we have finished all checks and our BBM is consistent.

From the structure of this BBM it is very clear that each row 
gives a $D_{k'}$ quiver and the different rows simply copies it $k$ times
according to the $Z_k$. This repetition hints that
there should be some kind of direct product, which is precisely what
we have.

\subsection{The Inverse Problem}
\label{subsec:inverse}
Now we address the inverse problem: given a BBM of type 
$Z$-$D$, with $k'$ 
vertical NS5-branes bounded by 2 ON$^0$-planes and $k$ 
horizontal NS5$'$-branes, what is the
corresponding orbifold, i.e., the group which acts on $\C^3$? 
The answer is now very clear:
if $k$ is odd we can choose the group $Z_{k} \times D_{k'}$ or 
$Z_{2k} \times D_{k'}$ with the action as defined in (\ref{gen_dir});
if $k$ is even, then we can choose the group to be 
$Z_{2k} \times D_{k'}$ with the same action.

In this above answer, we have two candidates when $k$ is odd since we recall
from discussions in \sref{subsec:BBMZD} the vertical direction of the BBM
for the group $Z_{2k}\times D_{k'}$
only has periodicity $\frac{k}{2}$ and the BBM separates into two pieces.
We must ask ourselves, what is the relation between these two 
candidates?
We notice that (\ref{gen_dir}) gives an non-faithful representation
of the group $Z_{2k}\times D_{k'}$. 
In fact, it defines a new group of which has the
faithful representation given by above matrix form 
and  is a factor group of
$Z_{2k} \times D_{k'}$ given by

\begin{equation}
\label{modH}
G:=(Z_{2k} \times D_{k'})/H,~~~{\rm with}~~~
	H=\langle1,\alpha^k \beta^{k'} \rangle
\end{equation}

In fact $G$ is isomorphic to $Z_{k} \times D_{k'}$. We can see this by
the following arguments.
denote the generators of $Z_{2k} \times D_{k'}$ as $\alpha,\beta,\gamma$ 
and those of $Z_{k} \times D_{k'}$ as 
$\tilde{\alpha},\tilde{\beta},\tilde{\gamma}$. 
An element of $G$ can be expressed as 
$[\alpha^a \beta^b \gamma^n] \equiv
[\alpha^{a+k} \beta^{b+k'} \gamma^n]$. We then see the homomorphism
from $G$ to $Z_{k} \times D_{k'}$ defined by
\[
[\alpha^a \beta^b \gamma^n] \longrightarrow 
\tilde{\alpha}^a \tilde{\beta}^{ak'+b} \tilde{\gamma}^n
\]
is in fact an isomorphism (we see that $[\alpha^a \beta^b \gamma^n]$ and
$[\alpha^{a+k} \beta^{b+k'} \gamma^n]$ are mapped to same element as
required; in proving this the $k=\odd$ condition is crucial).

We see therefore that given the data from the BBM, viz., $k$ and $k'$,
we can uniquely find the $\C^3$ orbifold singularity and our inverse
problem is well-defined.

\section{The General Twisted Case}
\label{sec:twist}
We have found \cite{Bo-Han} that the group $Z_{k} * D_{k'}$
(which in that paper we called $Z_{k} \times D_{k'}$) defined by
(\ref{gen1}) can be written in another form as 
$Z_k \rtimes D_{\frac{kk'}{\gcd(k,2k')}}$ where it becomes an
(internal) semidirect product.
We would like to know how the former, which is a special
case of what we shall call a
{\bf twisted} group\footnote{As mentioned in the
Nomenclature section, $*$ generically denotes twisted products
of groups.} is related to the direct
product example, which we shall call the {\bf untwisted} case,
upon which we expounded in the previous section.

The key relation which describes the semidirect product structure
was shown \cite{Bo-Han} to be
$\alpha \gamma =\beta^{\frac{2k'}{\gcd(k,2k')}} \gamma \alpha$.
This is highly suggestive and hints us to 
define a one-parameter family of groups\footnote{We note that
this is unambiguously the semi-direct product $\rtimes$: defining
the two subgroups $D := \langle \beta, \gamma \rangle$ and
$Z := \langle \alpha \rangle$, we see that $G(a) = DZ$ as cosets,
that $D \triangleleft G(a)$ and $D \cap Z = 1$, whereby all
the axioms of semi-directness are obeyed.}
$G(a) := \{Z_k \rtimes D_d\}$
whose presentations are

\begin{equation}
\label{gen_twist}
\alpha \beta=\beta \alpha,~~~~\alpha \gamma=\beta^a \gamma \alpha.
\end{equation}

When the parameter $a=0$, we have $G(0) = Z_k \times D_{k'}$ as discussed
extensively in the previous section. Also, when 
$a = \frac{kk'}{\gcd(k,2k')}$, $G(a)$
is the group $Z*D$ treated in the previous paper \cite{Bo-Han}. 
We are concerned with
members of $\{G(a)\}$ that satisfy the BBM conditions and though indeed
this family may not exhaust the list of all groups that satisfy
those conditions they do provide an illustrative subclass.

\subsection{Preserving the Irreps of $D_d$}
We see that the first of the BBM conditions is trivially
satisfied by our definition (\ref{gen_twist} of 
$G(a) := Z_k \rtimes D_d$. Therefore
we now move onto the second condition.
We propose that $G(a)$ preserves the 
structure of the irreps of the $D_d$ factor if $a$ is even.
The analysis had been given in detail \cite{Bo-Han} so
here we only review briefly.
Deducing from (\ref{gen_twist}) the relation, for $b \in \Z$,
\[
\alpha (\beta^b \gamma) \alpha^{-1}= \beta^{b+a} \gamma,
\]
we see that $\beta^b \gamma$ and $\beta^{b+a} \gamma$ belong to the same
conjugacy class after promoting $D_d$ to the semidirect product
$Z_k \rtimes D_d$. Now we recall from subsection \sref{subsec:Dk'} that
the conjugacy classes of $D_d$ are $\beta^0,\beta^d$,
$\beta^{\pm p}(p\neq 0,d)$, $\gamma \beta^{\even}$ and
$\gamma \beta^{\odd}$. Therefore we see that
when $a=\even$, the conjugacy structure of $D_d$ is
preserved since therein $\beta^b \gamma$ and $\beta^{b+a} \gamma$,
which we saw above belong to same conjugate class in $D_d$,
are also in the same conjugacy class in $G(a)$ and everything is fine.
However, when $a=\odd$, they live in two
different conjugacy classes at the level of $D_d$ but in the
same conjugacy class in $G(a)$ whence violating
the second condition. Therefore $a$ has to be even.

\subsection{The Three Dimensional Representation}
Now we come to the most important part of finding the 3-dimensional 
representations for $G(a)$, i.e., condition 3.
We start with the following form for the generators

\begin{equation}
\label{betagamma}
\beta = \left(  \begin{array}{ccc}
			1  & 0  & 0  \\
			0 & \omega_{2d} & 0  \\
			0 &  0 & \omega_{2d}^{-1}  
		\end{array}
	\right)
~~~~~~~
\gamma =\left(  \begin{array}{ccc} 
		1  &  0  &  0  \\
		0  &  0  &  i \\
		0  &  i &   0 
		\end{array}
	\right)
\end{equation}

and 

\begin{equation}
\label{alpha}
\alpha=\left( \begin{array}{ccc} 
\omega_{k}^{-(x+y)} &  0  & 0 \\
0 & \omega_{k}^{x} & 0  \\
0 & 0 & \omega_{k}^{y} 
		\end{array}
	\right)
\end{equation}
where $x,y \in \Z$ are yet undetermined integers
(notice that the form (\ref{alpha}) is fixed by the matrix
(\ref{betagamma}) of $\beta$ and the algebraic 
relation $\alpha \beta=\beta \alpha$). 
Using the defining relations (\ref{gen_twist}) of $G(a)$,
i.e relation $\alpha \gamma=\beta^a \gamma \alpha$, we 
immediately have the following constraint on $x$ and $y$:

\begin{equation}
\label{xy}
\omega_k ^{x-y} =\omega_{2d}^a
\end{equation}

which has integer solutions \footnote{Since (\ref{xy}) implies 
$\frac{2\pi (x-y)}{k} - \frac{2\pi a}{2d} = 2 \pi \Z$,
we are concerned with Diophantine equations of the form 
$\frac pq - \frac mn \in \Z$. This in turn requires that
$n p = m q \Rightarrow q = \frac{nl}{\gcd(m,n)},~l \in \Z$ by diving
through by the greatest common divisor of $m$ and $n$.
Upon back-substitution, we arrive at $p = \frac{m l}{\gcd(m,n)}$.}
only when

\begin{equation}
\label{ksol}
k=(\frac{2d}{\delta})l~~~~~l \in \Z~~{\rm and}~~\delta := \gcd (a,2d)
\end{equation}

with the actual solution being
\[
x-y=\frac{a}{\delta}l.
\]
Equation (\ref{ksol}) is a nontrivial condition which signifiess that for
arbitrary $k,2d,a$, the third of the BBM conditions may be violated,
and the solution, not found.
This shows that even though $G(a=\even)$
satisfies the first two of the BBM conditions, they can not
in general be applied to construct BBM's of Type $Z$-$D$ unless
(\ref{ksol}) is also respected. However, before starting
the general discussion of those cases of $Z*D$ where (\ref{ksol})
is satisfied, let us first see how the group treated before \cite{Bo-Han}
indeed satisfies this condition.

For $Z_k * D_{k'}$ \cite{Bo-Han} and defined by (\ref{gen1}),
let $\delta_1 := \gcd(k,2k')$. We have
$d=\frac{kk'}{\delta_1}$, $a=\frac{2k'}{\delta_1}$ from 
Proposition (3.1) in that
paper. Therefore $\delta=\gcd(a,2d)=a$ and
$k=\frac{2d}{\delta}$ so that (\ref{ksol}) is
satisfied with $l=1$ and we have the solution $x-y=1$.
Now if we choose $y=0$, then we have 

\begin{equation}
\label{alpha2}
\alpha=\left( \begin{array}{ccc} 
\omega_{k}^{-1} &  0  & 0 \\
0 & \omega_{k}^{1} & 0  \\
0 & 0 & 1 
		\end{array}
	\right).
\end{equation}

Combining with the matrices in (\ref{betagamma}), we see that they generate
a faithful 3-dimensional representation of $Z_k * D_{k'}$. It is easy to see
that what they generate is in fact isomorphic to a group with
matrix generators, as given in (\ref{gen_dir}),

\begin{equation}
\label{gen2}
\alpha^{-1} = \left(  \begin{array}{ccc}
			\omega_{2k}^{-2} & 0 & 0  \\
			 0 & \omega_{2k}^{1} & 0\\
			 0  &  0 & \omega_{2k}^{1}
		\end{array}
	\right)
~~~~~~~~
\beta = \left(  \begin{array}{ccc}
			1  & 0  & 0  \\
			0 & \omega_{2d} & 0  \\
			0 &  0 & \omega_{2d}^{-1}  
		\end{array}
	\right)
~~~~~~~
\gamma =\left(  \begin{array}{ccc} 
		1  &  0  &  0  \\
		0  &  0  &  i \\
		0  &  i &   0 
		\end{array}
	\right)
\end{equation}

by noticing that $\alpha^{-1} \beta^{\frac{k'}{\delta}}$ in
(\ref{gen2}) is precisely (\ref{alpha2}).
But this is simply a non-faithful representation of 
$Z_{2k} \times D_{d=\frac{kk'}{\gcd(k,2k')}}$, our direct
product example! 
Furthermore, when $k=odd$,
by recalling the results of \sref{subsec:inverse} we conclude
in fact that the group $Z_k * D_{k'}$
is isomorphic to $Z_k \times D_d$.
However, for $k=\even$, although $Z_k * D_{k'}$ is 
still embeddable into $Z_{2k} \times D_{d=\frac{kk'}{\gcd(k,2k')}}$
with a non-faithful representation (\ref{gen_dir}), it is 
not isomorphic to $Z_k \times D_d$ and the BBM thereof
corresponds to an intrintically twisted case (and unlike when
$k=\odd$ where it is actually isomorphic to a direct product group). 
We emphasize here an obvious but crucial fact exemplified by
(\ref{modH}): {\it non-faithful representations of a group $A$
can be considered as the faithful representation of a new
group $B$ obtained by quotienting an appropriate normal subgroup
of $A$.}
This is what is happening above.
This explains also why we have succeeded \cite{Bo-Han} in 
constructing the BBM only when we wrote $Z_k * D_{k'}$ in the form 
$Z_{k} \rtimes D_{d=\frac{kk'}{\gcd(k,2k')}}$.

Now let us discuss the general case. We recall from the
previous subsection that $a$ has to be even; we thus let $a:=2m$.
With this definition, putting (\ref{xy}) into (\ref{alpha},)
we obtain for the quantity $\alpha \beta^{-m}$:

\begin{equation}
\label{atilde}
\tilde{\alpha}=\alpha \beta^{-m}=\left( \begin{array}{ccc} 
\omega_{k}^{-2y}\omega_{2d}^{2m} &  0  & 0 \\
0 & \omega_{k}^{y}\omega_{2d}^{-m} & 0  \\
0 & 0 & \omega_{k}^{y} \omega_{2d}^{-m}
		\end{array}
	\right)
\end{equation}

This $\tilde{\alpha}$ generates a cyclic group $Z_{\tilde{k}}$ 
and combined with (\ref{betagamma}) gives the direct 
product group of $Z_{\tilde{k}} \times D_d$, but
with a non-faithful representation as in (\ref{gen_dir}).
Therefore for the general twisted case, we can obtain 
the BBM of $Z$-$D$ type of $G(a)$ by imbedding $G(a)$ into a 
larger group $Z_{\tilde{k}} \times D_d$
which is a direct product 
just like we did  for $Z_k * D_{k'}$ embeding to
$Z_{k} \rtimes D_{d=\frac{kk'}{\gcd(k,2k')}}$ two
paragraphs before,  and for which, by our
etude in \sref{sec:dir}, a consistent BBM can always be
established.
However, we need to emphasize that in general such an
embedding (\ref{atilde}) gives non-faithful representations 
so that the quiver diagram of the twisted group
will be a union of disconnected pieces, as demanded by
Proposition \ref{nonfaith}, each of which
corresponds to a Type $Z$-$D$ BBM.
We summarise these results by stating
\begin{proposition}
\label{embed}
The group $G(a):=Z_k*D_d$ satisfies the BBM conditions if $a$ is even and
the relation (\ref{ksol}) is obeyed. In this case its matrices
actually furnish a non-faithful representation of a direct
product $\tilde{G} := Z_{\tilde{k}} \times D_d$ and hence 
affords a BBM\footnote{Though possibly disconnected with
the number of components 
depending on the order of an Abelian subgroup $H\triangleleft \tilde{G}$.}
of Type $Z$-$D$.
\end{proposition}
This action of $G(a) \hookrightarrow \tilde{G}$ is what we mean by embedding.
We conclude by saying that the simple example of \sref{sec:dir}
where the BBM is constructed for untwisted (direct-product)
groups is in fact general and Type $Z$-$D$ BBM's can be
obtained for twisted groups by imbedding into such direct-product
structures.

\section{A New Class of $SU(3)$ Quivers}
\label{sec:new}
It would be nice to see whether the ideas presented in the above sections can be 
generalised to give the BBM of other types such as Type $Z$-$E$,
$Z$-$d$ or $D$-$E$ 
whose definitions are obvious. Moreover, $E$ refers to the exceptional
groups $\widehat{E_{6,7,8}}$ and $d$ the ordinary dihedral group.
Indeed, we must first have the brane setups for these groups.
Unfortunately as of yet the $E$ groups still remain elusive.
However we will give an account of the
ordinary dihedral groups and the quiver theory thereof, as well as
the ordinary $E$ groups from a new perspectively from an earlier 
work \cite{Han-He}.
These shall give us a new class of $SU(3)$ quivers.

We note that, as pointed out \cite{Han-He}, the ordinary
di-, tetra-, octa- and iscosa-hedral groups (or $d$, $E_6,7,8$ 
respectively) are excluded from the
classification of the discrete finite subgroups of $SU(2)$ because
they in fact belong to the centre-modded group $SO(3) \cong SU(2)/\Z_2$.
However due to the obvious embedding $SO(3) \hookrightarrow SU(3)$,
these are all actually $SU(3)$ subgroups. Now the $d$-groups were not
discussed before \cite{Han-He} because they did not have non-trivial
3-dimensional irreps and are not considered as non-trivial (i.e.,
they are fundamentally 2-dimensional collineation groups) in the
standard classification of $SU(3)$ subgroups; or in a mathemtical language
\cite{Trans}, they are transitives. Moreover, $E_6$ is precisely
what was called $\Delta(3 \times 2^2)$ earlier \cite{Han-He}, $E_7$, 
$\Delta(6 \times 2^2)$ and $E_8$, $\Sigma_{60}$ and were discussed
there. However we shall here see all these groups together under a new
light, especially the ordinary dihedral group to which we now turn.

\subsection{The Group $d_{k'}$}
The group is defined as
\[
\beta^{k'} = \gamma^2 = 1,~~~~~~\beta \gamma=\gamma \beta^{-1},
\]
and differs from its binary cousin $D_{k'}$ in subsection
\sref{subsec:Dk'} only by having the orders of $\beta, \gamma$
being one half of the latter. Indeed, defining the normal
subgroup $H := \{1,\beta^{k'}\} \triangleleft D_{k'}$ we have
\[
d_{k'} \cong D_{k'}/H.
\]
We can subsequently obtain the character table of $d_{k'}$ from
that of $D_{k'}$ by using the theory of subduced representations,
or simply by keeping all the irreps of $D_{k'}$ which are 
invariant under the equivalence by $H$.
The action of $H$ depends on the parity of $k'$. When
it is even, the two conjugacy classes $(\gamma \beta^{even})$ and 
$(\gamma \beta^{odd})$ remain separate. Furthermore, 
the four 1-dimensional irreps are invariant while for the
2-dimensionals we must restrict the index $l$ as defined in
subsection \sref{subsec:Dk'} to $l=2,4,6,...,k'-2$ so as to
observe the fact that the two conjugacy classes 
$\{\beta^a,\beta^{-a}\}$ and $\{\beta^{k-a},\beta^{a-k}\}$
combine into a single one. All in all, we have 4 1-dimensional 
irreps and $\frac{k'-2}{2}$ 2-dimensionals.
On the other hand, for $k'$ odd, we have the two classes
$(\gamma \beta^{even})$ and $(\gamma \beta^{odd})$ collapsing
into a single one, whereby we can only keep $\chi^1,\chi^3$
in the 1-dimensionals and restrict $l=2,4,6,...,k'-1$ for the
2-dimensionals. Here we have a total of 2 1-dimensional
irreps and $\frac{k'-1}{2}$ 2-dimensionals.

In summary then, the character tables are as follows:
\[
\begin{array}{ll}
\begin{array}{|c|c|c|c|c|c|c|}
\hline
        & 1 & 2 & 2 & \cdots  & 2 & n \\ \hline
\Gamma_{1} & 1 & 1 & 1 & \cdots  & 1 & 1 \\ \hline
\Gamma_{2} & 1 & 1 & 1 & \cdots  & 1 & -1 \\ \hline
\Gamma_{3} & 2 & 2\cos \phi  & 2\cos 2\phi  & \cdots  & 2\cos m\phi  & 0 \\ \hline
\Gamma_{4} & 2 & 2\cos 2\phi  & 2\cos 4\phi  & \cdots  & 2\cos 2m\phi  & 0 \\ \hline
        \vdots  & \vdots  & \vdots  & \vdots  & \cdots  & \vdots  & \vdots  \\ \hline
\Gamma_{\frac{k'+3}{2}} & 2 & 2\cos m\phi  & 2\cos 2m\phi  & \cdots 
        & 2\cos m^{2}\phi  & 0 \\ \hline
\end{array}
& 
\begin{array}{l}
k' $ odd$  \\ 
m=\frac{k'-1}{2} \\ 
\phi =\frac{2\pi }{k'}
\end{array}
\end{array}
\]

\[
\begin{array}{ll}
\begin{array}{|c|c|c|c|c|c|c|c|c|}
\hline
        & 1 & 2 & 2 & \cdots  & 2 & 1 & n/2 & n/2 \\ \hline
\Gamma_{1} & 1 & 1 & 1 & \cdots  & 1 & 1 & 1 & 1 \\ \hline
\Gamma_{2} & 1 & 1 & 1 & \cdots  & 1 & 1 & -1 & -1 \\ \hline
\Gamma_{3} & 1 & -1 & 1 & \cdots  & (-1)^{m-1} & (-1)^{m} & 1 & -1 \\ \hline
\Gamma_{4} & 1 & -1 & 1 & \cdots  & (-1)^{m-1} & (-1)^{m} & -1& 1 \\ \hline
\Gamma_{5} & 2 & 2 \cos \phi  & 2\cos 2\phi  & \cdots  & 
        2\cos(m-1)\phi  & 2\cos m\phi  & 0 & 0 \\ \hline
\Gamma_{6} & 2 & 2\cos 2\phi  & 2\cos 4\phi  & \cdots  & 2\cos
        2(m-1)\phi  & 2\cos 2m\phi  & 0 & 0 \\ \hline
\vdots  &\vdots  &\vdots  &\vdots  &\cdots  &\vdots  &\vdots  &\vdots  
	&\vdots \\ \hline
\Gamma_{\frac{k'+6}{2}} & 2 & 2\cos (m-1)\phi  & 2\cos 2(m-1)\phi  &
        \cdots  & 2\cos (m-1)^{2}\phi  & 2\cos m(m-1)\phi  & 0 & 0 \\ \hline
\end{array}
&
\begin{array}{l}
k' $ even$  \\ 
m=\frac{k'}{2} \\ 
\phi =\frac{2\pi }{k'}
\end{array}
\end{array}
\]

\subsection{A New Set of Quivers}
\begin{figure}
\centerline{\psfig{figure=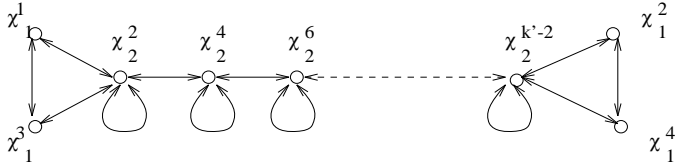,width=3.5in}}
\caption{The quiver diagram for $d_{k=\even}$. Here the notation of the irreps
	placed on the nodes is
 	borrowed from $D_k$ in \sref{subsec:Dk'}. Notice that it
	gives a finite theory with non-chiral matter content.}
\label{fig:dkeven}
\end{figure}

\begin{figure}
\centerline{\psfig{figure=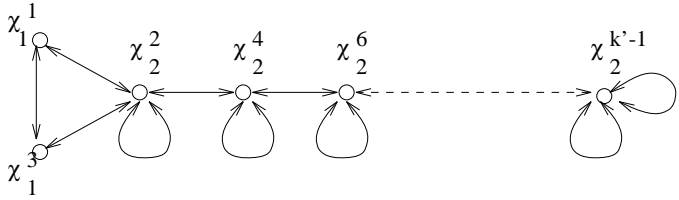,width=3.5in}}
\caption{The quiver diagram for $d_{k=\odd}$. Here again we use the notation 
	of the irreps of $D_k$ to index the nodes. 
	Notice that the theory is again finite and
	non-chiral.}
\label{fig:dkodd}
\end{figure}

Now we must choose an appropriate $SU(3)$ decomposition of the
{\bf 3} for our group in order to make physical sense for
the bifundamentals. The choice is
\[
{\bf 3} \longrightarrow \chi_1^3+\chi_2^2.
\]
Here, we borrow the notation of the irreps of $d_k$ from 
$D_k$ in \sref{subsec:Dk'}. 
The relationship between the irreps of the
two is discussed in the previous subsection. The advantage of using this
notation is that we can readily use the tabulated tensor decompositions
of $D_k$ in \sref{subsec:Dk'}.
With this chosen decomposition, we can immediately arrive at 
the matter matrices $a_{ij}$ and subsequent quiver diagrams.
The $k' = \even$ case gives a quiver which is very much like the
affine $\widehat{D_{k'+2}}$ Dynkin
Diagram, differing only at the two ends, where the nodes corresponding
to the 1-dimensionals are joined, as well as the existence of
self-joined nodes.
This is of course almost what one would expect from an ${\cal N}=2$
theory obtained from the binary dihedral group as a finite
subgroup of $SU(2)$; this clearly reflects the intimate relationship
between the ordinary and binary dihedral groups. 
The quiver is shown in \fref{fig:dkeven}.
On the other hand, for $k'$ odd, we have a quiver which looks like
an ordinary $D_{k'+1}$ Dynkin Diagram with 1 extra line
joining the 1-nodes as well as self-adjoints.
This issue of the dichotomous appearance of affine
and ordinary Dynkin graphs of the D-series in brane setups
has been raised before \cite{Han-Zaf2,Bo-Han}. The diagram for $k'$ odd
is shown in \fref{fig:dkodd}.

\begin{figure}
\centerline{\psfig{figure=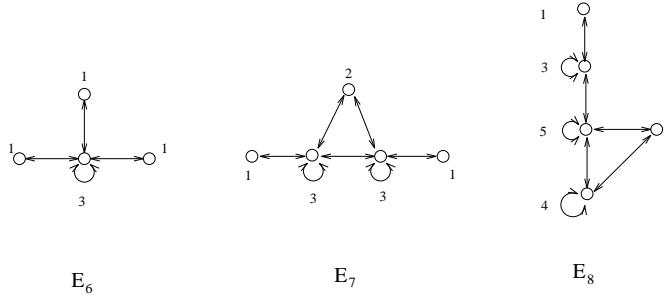,width=3.5in}}
\caption{The quiver diagrams for
	$E_6 = A_4 = \Delta(3 \times 2^2)$,
	$E_7 = S_4 = \Delta(6 \times 2^2)$ and
	$E_8 = A_5 = \Sigma_{60}$. The theories are
	finite and non-chiral.}
\label{fig:excep}
\end{figure}

For completeness and comparison we hereby also include
the 3 exceptional groups of $SO(3) \subset SU(3)$. For these,
we must choose the {\bf 3} to be the unique (up to 
automorphism among the conjugacy classes) 3-dimensional irrep.
Any other decompostion leads to non-faithful representations of
the action and subsequently, by our rule discussed earlier,
to disconnected quivers. This is why when they were considered
as $SU(2)/\Z_2$ groups with 
${\bf 3} \rightarrow {\bf 1} \oplus {\bf 2}$ chosen, 
uninteresting and disconnected quivers were obtained
\cite{Han-He}. Now under this new light, we present
the quivers for these 3 groups in \fref{fig:excep}.

There are two points worth emphasising. All the above quivers
correspond to theories which are finite and non-chiral.
By {\bf finite} we mean the condition \cite{Law-Vafa}
for anomaly cancelation, that the matter matrix $a_{ij}^R$ must
satisfy
\[
\sum\limits_j a_{ij}^R \dim(r_j) = \sum\limits_j a_{ji}^R \dim(r_j)
\]
What this mean graphically is that for each node, the sum of the
indices of all the neighbouring nodes flowing thereto (i.e., having
arrows pointing to it) must equal to the sum of those flowing
therefrom, and must in fact, for an ${\cal N} = 1$ theory, be
equal to 3 times the index for the node itself.
We observe that this condition is satisfied for all the quivers
presented in \fref{fig:dkodd} to \fref{fig:excep}.

On the other hand by {\bf non-chiral} we mean that for every
bi-fundamental chiral multiplet $(N_i,\bar{N_j})$ there exists
a companion $(N_j,\bar{N_i})$ (such that the two combine together to give
a bi-fundamental hypermultiplet in the sense of ${\cal N}=2$). 
Graphically, this dictates that
for each arrow between two nodes there exists another in the
opposite direction, i.e., the quiver graph is unoriented.
Strangely enough, non-chiral matter content is a trademark
for ${\cal N}=2$ theories, obtained from 
$\C^2 / \Gamma \subset SU(2)$ singularities, while ${\cal N}=1$
usually affords chiral (i.e., oriented quivers) theories.
We have thus arrived at a class of finite, non-chiral
${\cal N} = 1$ super Yang-Mills theories. 
This is not that peculiar because all these groups belong to $SO(3)$ and
thus have real representations; the reality compel the existence
of complex conjugate pairs.
The more interesting fact is that these groups give quivers that
are in some sense in between the generic non-chiral
$SU(2)$ and chiral $SU(3)$ quiver theories. Therefore
we expect that the corresposnding gauge theory will have better properties,
or have more control, under the evolution along some energy scale.

\subsection{An Interesting Observation}
Having obtained a new quiver, for the group $d_k$, it is natural to ask 
what is the corresponding brane setup. 
Furthermore, if we can realize such a brane setup,
can we apply the ideas in the previous sections to realize the
BBM of Type $Z$-$d$? We regrettably have no answers at this stage as
attempts at the brane setup have met great difficulty.
We do, however, have an
interesting brane configuration which gives the correct matter content of 
$d_k$ but has a different superpotential.
The subtle point is that $d_k$ gives only ${\cal N}=1$ 
supersymmetry and unlike ${\cal N}=2$, one must specify both the matter 
content and the superpotential. Two theories with the same matter content but 
different superpotential usually have different low-energy behavior.

\begin{figure}
\centerline{\psfig{figure=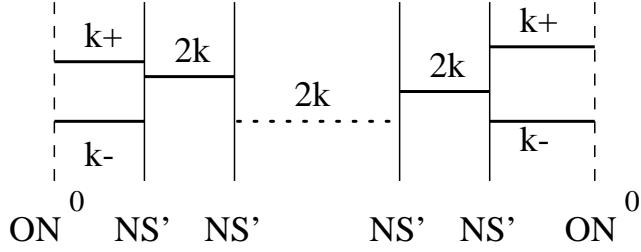,width=3.3in}}
\caption{The brane configuration which gives the same matter content as 
	the $d_{k=\even}$ quiver.}
\label{fig:d_k}
\end{figure}

We now discuss the brane configuration connected with
$d_k$, which turns out to be
a rotated version of the configuration for $D_k$ as given by 
Kapustin \cite{Kapustin}
(related examples \cite{Bo-Han,Erlich} on how rotating branes
breaks supersymmetry further may be found).
In particular we rotate all NS5-branes (along direction
(12345)) between the two ON$^0$-plane as drawn in Figure 1 of Kapustin
\cite{Kapustin} 
to NS5$'$-branes (along direction (12389)). The setup is shown in
\fref{fig:d_k}.
Let us analyse this brane setup more carefully. 
First when we end D4-branes (extended along direction
(1236)) on the ON$^0$-plane, they can have two different charges: positive 
or negtive. With the definition of the matrix
\[
\Omega=\left(  \begin{array}{cc}
	1_{k+\times k+}  &  0  \\
	0  &  -1_{k-\times k-}
	\end{array}  \right),
\]
the projection on the Chan-Paton matrix of the D4-branes is as follows.
The scalar fields in the D4-worldvolume are projected as
\[
\phi^{\alpha}= \Omega \phi^{\alpha}\Omega^{-1}
~~~{\rm and}~~~
\phi^{i}=- \Omega \phi^{i}\Omega^{-1}
\]
where $\alpha$ runs from $4$ to $5$ and describes the oscillations of the
D4-branes in the directions parallel to the ON$^0$-plane while $i$ runs from
$7$ to $9$ and describes the transverse oscillations.
If we write the scalars as matrice in block form, 
the remaining scalars that survive the projection are
\[
\phi^{\alpha}=\left(  \begin{array}{cc}
	U_{k+\times k+}  &  0  \\
	0  &  U_{k-\times k-} \end{array} \right)
~~~{\rm and}~~~
\phi^{i}=\left(   \begin{array}{cc}
	0  &  U_{k+ \times k-}  \\
	U_{k-\times k+}  &  0  
\end{array}  \right).
\]

\begin{figure}
\centerline{\psfig{figure=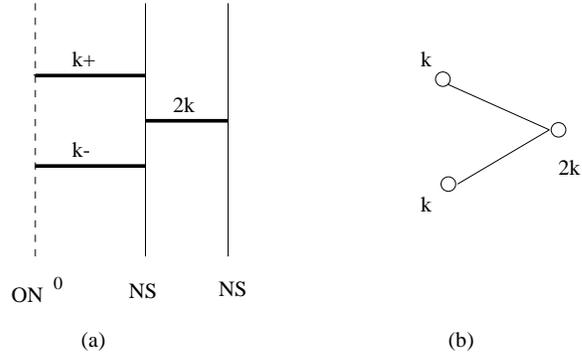,width=3.0in}}
\caption{(a). The brane configuration of the projection using NS5-branes.
	(b). The quiver diagram for the brane configuration in (a).}
\label{fig:NSp}
\end{figure}

From these we immediately see that $\phi^{\alpha}$ give scalars
in the adjoint representation and $\phi^{i}$, in the bifundamental representation.
Next we consider the projection conditions when we end the other side of 
our D4-brane on the NS-brane.
If we choose the NS5-brane to extend along (12345), then the scalars $\phi^{\alpha}$
will be kept while $\phi^{i}$ will be projected out and we would have an
${\cal N}=2$ $D_k$ quiver (see \fref{fig:NSp}).

\begin{figure}
\centerline{\psfig{figure=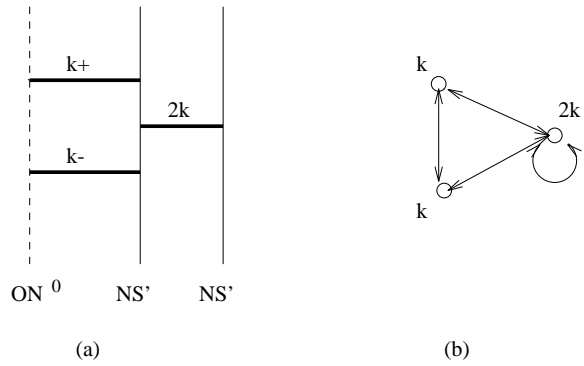,width=3.0in}}
\caption{(a). The brane configuration of projection using NS5$'$-branes.
	(b). The quiver diagram for the brane configuration in (a).}
\label{fig:NSrp}
\end{figure}

However, if we choose the NS5-branes to extend along (12389), then
$\phi^{\alpha}$ and $\phi^{i=7}$ will be projected out while
$\phi^{i=8,9}$ will be kept. It is in this case that
we see immediately that we obtain the same matter content 
as one would have from a $d_{k=\even}$ orbifold discussed in the
previous subsection (see \fref{fig:NSrp}).

Now we explain why the above brane configuration, though giving the
same matter content as
the $d_{k=\even}$,
is insufficient to describe the full theory.
The setup in \fref{fig:d_k} is obtained by the
rotation of NS-branes to NS$'$-branes; in this process of rotation, 
in general we change the geometry from an orbifold to a conifold. 
In other words, by rotating,
we break the ${\cal N}=2$ theory to ${\cal N}=1$ by giving masses to 
scalars in the ${\cal N}=2$ vector-multiplet. 
After integrating out the massive adjoint scalar in low
energies, we usually get quartic terms in the superpotential (for more 
detailed discussion of rotation see Erlich et al. \cite{Erlich}). 
Indeed Klebanov and Witten \cite{Kle-Wit}
have explained this point carefully and shows that the quartic terms will exist
even at the limiting case when the angle of rotation is $\frac{\pi}{2}$ and the
NS5-branes become NS5$'$-branes. On the other hand, the superpotential for
the orbifold singularity of $d_k$ contains only cubic terms as required by
Lawrence et al. \cite{Law-Vafa} and as we emphasized in \sref{sec:dir}.
It still remains an intersting problem to construct consistent brane
setups for $d_k$ that also has the right superpotential; this would
give us one further stride toward attacking non-Abelian brane configurations.

\section{Conclusions and Prospects}
\label{sec:con}
As inspired by the Brane Box Model (BBM) constructions \cite{Bo-Han}
for the group $Z_k * D_{k'}$ generated by (\ref{gen1}), we have discussed
in this paper a class of groups which are generalisations thereof.
These groups we have called the twisted groups 
(that satisfy BBM conditions).
In particular we have analysed at great length, the simplest memeber of this
class, namely the direct product $Z_k \times D_d$, focusing on how the
quiver theory, the BBM construction as well as the inverse problem (of
recovering the group by reading the brane setup) may be established.
The brane configuration for such an example, as in \fref{fig:BBM1}, 
we have called a {\bf BBM of Type $Z$-$D$}; consisting generically of a
grid of NS5-branes with the horizontal direction bounded by 2 ON-planes
and the vertical direction periodically identified.
We have also addressed, as given in Proposition \ref{nonfaith} the
issue of how non-faithful representations lead to disconnected
quivers graphs, or in other words several disjunct pieces of the
BBM setup.

What is remarkable is that the twisted groups, of which the one 
in our recent paper \cite{Bo-Han} 
is a special case, can under certain circumstances
be embedded into a direct product structure (by actually furnishing
a non-faithful representation thereof). This makes our na\"{\i}ve
example of $Z_k \times D_d$ actually possess great generality as
the twisted cases untwist themselve by embedding into this, in a sense,
universal cover in the fashion of Proposition \ref{embed}.
What we hope is that this technique may be extended to address
more non-Abelian singularities of $\C^3$, whereby the generic finite 
discrete group $G \subset SU(3)$
maybe untwisted into a direct-product cover. In order to do so,
it seems that $G$ needs to obey a set of what we call {\bf BBM conditions}.
We state these in a daring generality: (1) That $G$ maybe written as
a semi-direct product $A \rtimes B$, (2) that the structure of the irreps
of $G$ preserves those of the factors $A$ and $B$ and (3) that there
exists a decomposition into the irreps of $G$ consistent with the unitarity
and determinant 1 constraints of $SU(3)$.

Indeed it is projected and hoped, with reserved optimism,
that if $A,B$ are $SU(2)$ subgroups for
which a brane setup is known, the techniques presented above may inductively
promote the setup to a BBM (or perhaps even brane cube for $SU(4)$ 
singularities). Bearing this in mind, we proceeded further to study
more examples, hoping to attack for example, BBM's of the $Z$-$d$
type where $d$ is the ordinary dihedral group. Therefrom arose our interest
in the ordinary groups $d,E_{6,7,8}$ as finite subgroups of
$SO(3) \subset SU(3)$. These gave us a new class of quiver
theories which have ${\cal N}=1$ but non-chiral matter content.
Brane setups that reproduce the matter content, but unfortunately not the
superpotential, have been established for the ordinary dihedral groups.
These give an interesting brane configuration involving rotating NS5-brane
with respect to ON-planes.

Of course much work remains to be done. In addition to finding the
complete brane setups that reproduce the ordinary dihedral quiver as well as
superpotential, we have yet to clarify the BBM conditions for groups
in general and head toward that beacon of brane realisations of non-Abelian
orbifold theories.

\section*{Acknowledgements}
We would like to extend our sincere gratitute to A. Uranga for 
valuable discussions and comments, N. Moeller and J. S. Song for
suggestions and help. Indeed we are further indebted to the CTP and LNS
of MIT for their gracious patronage and YHH, to the NSF for her
generous support. A. H. is supported in part by an A. P. Sloan foundation fellowship
and by a DOE OJI Award.



\begin{thebibliography}{9}
\bibitem{Han-Wit} A. Hanany and E. Witten, ``Type IIB Superstrings,
	BPS monopoles, and 
        Three-Dimensional Gauge Dynamics,'' hep-th/9611230.

\bibitem{Giveon} A. Giveon and D. Kutasov, ``Brane Dynamics and Gauge Theory,''
        hep-th/9802067.

\bibitem{IS} K. Intriligator, N. Seiberg, ``Mirror Symmetry in Three 
	Dimensional Gauge Theories,'' Phys.Lett. B387 (1996) 513-519,
	hep-th/9607207.

\bibitem{P-Zaf}  M. Porrati, A. Zaffaroni, ``M-Theory Origin of 
	Mirror Symmetry	 in Three Dimensional Gauge Theories'',
	Nucl.Phys. B490 (1997) 107-120,
	hep-th/9611201.

\bibitem{Boer} Jan de Boer, Kentaro Hori, Hirosi Ooguri, Yaron Oz and Zheng Yin,
        ``Mirror Symmetry in Three-dimensional Gauge Theories, $SL(2,Z)$ and
        D-Brane Moduli Spaces,'' hep-th/9612131.

\bibitem{Kapustin} A. Kapustin, ``$D_n$ Quivers from Branes,'' hep-th/9806238.

\bibitem{Elitzur}  S. Elitzur, A. Giveon and D. Kutasov, ``Branes and N=1 
	Duality in String Theory,'' Phys.Lett. B400 (1997) 269-274, hep-th/9702014.\\
	S. Elitzur, A. Giveon, D. Kutasov, E. Rabinovici and A. Schwimmer,
        ``Brane dynamics and $N=1$ supersymmetric gauge theory,'' Nucl.Phys.B 
        {\bf 505}(1997) 202-250.

\bibitem{Mlift} E. Witten, ``Solutions of Four-Dimensional Field Theories Via M Theory,''
        hep-th/9703166.

\bibitem{Karl}  Karl Landsteiner, Esperanza Lopez, David A. Lowe, `` N=2 Supersymmetric 
	Gauge Theories, Branes and Orientifolds'',
	Nucl.Phys. B507 (1997) 197-226. hep-th/9705199. \\
	 Karl Landsteiner, Esperanza Lopez, ``New Curves from Branes'',
	Nucl.Phys. B516 (1998) 273-296, hep-th/9708118.

\bibitem{Quiver} M. Douglas and G. Moore, ``D-Branes, Quivers, and ALE Instantons,''
        hep-th/9603167.

\bibitem{Johnson}  Clifford V. Johnson, Robert C. Myers, "Aspects of Type IIB 
	Theory on ALE Spaces",  Phys.Rev. D55 (1997) 6382-6393.hep-th/9610140.

\bibitem{Law-Vafa}
        A. Lawrence, N. Nekrasov and C. Vafa, ``On Conformal Field
        Theories in Four Dimensions,'' hep-th/9803015.

\bibitem{Han-Zaf} A. Hanany and A. Zaffaroni, ``On the Realisation of
	Chiral Four-Dimenaional Gauge Theories Using Branes,'' hep-th/9801134.

\bibitem{Han-Ura} A. Hanany and A. Uranga, ``Brane Boxes and Branes on Singularities,''
        hep-th/9805139.

\bibitem{Han-S} Amihay Hanany, Matthew J. Strassler and Angel M. Uranga, 
        ``Finite Theories and Marginal Operators on the Brane,''
        hep-th/9803086, JHEP 9806 (1998) 011.

\bibitem{Bo-Han} Bo Feng, Amihay Hanany, Yang-Hui He, ``The $Z_k \times 
	D_{k'}$ Brane Box Model,'' hep-th/9906031.

\bibitem{Karch1} Andreas Karch, Dieter Lust, Douglas J. Smith,
	``Equivalence of Geometric Engineering and Hanany-Witten
	via Fractional Branes'', Nucl.Phys. B533 (1998) 348-372, hep-th/9803232.

\bibitem{Han-Zaf2}  A. Hanany and A. Zaffaroni, ``Issues on Orientifolds: On the Brane
        Construction of Gauge Theories with $SO(2n)$ Global Symmetry,'' hep-th/9903242.

\bibitem{Han-He} A. Hanany and Y.-H. He, ``Non-Abelian Finite Gauge Theories,''
        hep-th/9811183, JHEP 9902 (1999) 013.

\bibitem{Trans} A. Hanany and Y.-H. He, ``A Monograph on the 
        Classification of the Discrete Subgroups of $SU(4)$,'' hep-th/9905212.\\
	S.-T. Yau and Y. Yu, ``Gorenstein Quotients Singularities in
        Dimension Three,'' Memoirs of the AMS, 505, 1993.


\bibitem{Erlich} J. Erlich, A. Hanany, and A. Naqvi, ``Marginal Deformations 
        from Branes,'' hep-th/9902118.


\bibitem{Kle-Wit} Igor R. Klebanov, Edward Witten, ``Superconformal Field Theory 
	on Threebranes at a Calabi-Yau Singularity'',
 	Nucl.Phys. B536 (1998) 199-218,hep-th/9807080.
\end{thebibliography}
\end{document}